# A Framework for a Comprehensive National Future Readiness Index

Ali Qassim Jawad, Royal Academy of Management, Oman

Xavier Sala-i-Martin, Columbia University, NBER and Royal Academy of Management

August 2025

# Abstract

This paper introduces the Index of Future Readiness (IFR), a novel framework for assessing a country's capacity to withstand, adapt to, and prosper within an environment of continuous and accelerating change. The framework builds on the classical distinction, first emphasized in economic discourse by Robert E. Lucas Jr. in the 1970s, between temporary shocks, which call for economic resilience, and permanent shocks, which require adaptive capacity. The IFR applies a whole-of-society perspective, drawing on Ali Qassim Jawad's GBC model, to evaluate the preparedness of three central actors (government, businesses, and citizens) across six foundational dimensions. Unlike conventional benchmarking instruments, the IFR explicitly integrates both resilience, understood as the ability to "bounce back" from temporary shocks, and adaptive capacity, conceived as the ability to "bounce forward" in the face of structural or permanent disturbances. By enabling both sectoral and thematic analysis, the IFR provides policymakers with actionable insights and thus serves as a dynamic, policy-relevant tool for guiding national development in an era of systemic volatility.

# 1. Introduction

The 21st century is characterized by an unprecedented pace and complexity of global change, demanding a re-evaluation of how nations measure and cultivate their capacity for long-term prosperity. The contemporary global landscape is defined by an escalating array of interconnected challenges. The World Economic Forum's Global Risks Report 2025 highlights an increasingly fractured global environment, marked by intensifying geopolitical, environmental, societal, and technological challenges that threaten stability and progress. This report underscores a pervasive sense of declining optimism, with a significant majority of experts anticipating an "unsettled" or "turbulent" global outlook over the next few years and decades (World Economic Forum (2025).)

Environmental challenges, particularly climate change, represent a dominant and escalating risk. The increasing frequency of extreme weather events directly threatens local and global economies. The economic impacts are profound, including negative effects on productivity, business investment, and critical infrastructure such as housing and transportation networks. Key sectors like agriculture face significant damages, with projections of substantial crop yield reductions due to increased heat, drought, and extreme rainfall. These long-term economic effects are widely expected to be increasingly negative and widespread, posing catastrophic risks for some regions (World Economic Forum (2025).).

Concurrently, rapid technological disruptions are fundamentally reshaping economic systems. Advancements in areas such as generative artificial intelligence, automation, and the Internet of Things are creating and destroying business models, supply chains, and employment patterns at an unprecedented pace. These disruptive forces have the potential to impact growth, employment, and inequality, leading to large-scale displacement of labor that necessitates significant re-skilling and adjustments in workforce development strategies. The shift from fixed capital assets to technology-based firms as the largest global companies illustrates the profound transformation underway.

Geopolitical shifts further amplify this volatility. Escalating geopolitical tensions contribute to economic slowdowns, increasing national debt burdens, and the fragmentation of global trade. The weaponization of economic interdependence, manifested through trade wars and sanctions, disrupts global supply chains and erodes trust among nations. These dynamics create a "fragmented world economy" that is not a temporary detour but a "new context for competition," directly impacting financial markets and stock prices.

Beyond these macro-level trends, pervasive societal pressures also contribute to instability. Societal polarization, growing inequality, and involuntary migration are identified as significant risks, indicating fragile social stability (World Economic Forum (2025).).

Over the past several decades, governments across the world have pursued two principal macroeconomic policy objectives: the stabilization of economic fluctuations and the promotion of sustained economic growth. But to deal with the unprecedented risks and uncertainty of today's world, we need to fundamentally shift national strategic thinking, moving beyond the stabilization of economic fluctuations and the maximization of economic growth. We need to start thinking about "Future Readiness".

The strategic management literature conceptualizes future readiness as "the capacity of firms to anticipate and adapt to external changes through a balanced integration of strategic foresight and tactical adaptability" (see the "Future Readiness Indicator" developed by the IMD (2025). See also Yu et al. (2022)). Building on this foundation, the present paper seeks to extend the concept beyond the corporate domain, applying it to the macroeconomic level. In doing so, the analysis shifts the focus from firms to entire national economies, with the objective of assessing how countries, rather than individual organizations, can cultivate and sustain future readiness.

Benchmarking indexes play a crucial role in international economic and development discourse by providing comparative assessments of national performance across various dimensions. These tools offer significant utility for policymakers, researchers, and international organizations. They inform a myriad of institutional actions, helping countries identify areas of favorable or unfavorable performance relative to peers. By systematically comparing and learning from best practices, benchmarking enables the discovery of new ideas to enhance achievement and supports strategic planning capabilities. Furthermore, the public nature of these results promotes an environment of accountability and transparency, extending to governing agencies, students, and other stakeholders. Benchmarks are integral to quality assurance and improvement processes, allowing for the assessment of performance against specified targets or standards and monitoring outcomes at both local and competitor levels.

For example, the World Economic Forum's Global Competitiveness Index (GCI) was published annually between 2004 (Artadi and Sala-i-Martin (2004)) and 2019 and assessed the ability of countries to provide high levels of prosperity to their citizens, which is fundamentally linked to how productively a country uses its available resources. It measures the set of institutions, policies, and factors that determine sustainable current and medium-term economic prosperity. The GCI comprised over 110 variables organized into twelve pillars (e.g., institutions, infrastructure, macroeconomic framework, health, education, goods/labor/financial markets, technology, innovation).

Along the same lines, the International Institute for Management or IMD publishes the World Competitiveness Ranking (WCR) since 1989. The WCR benchmarks the performance of 63 countries using 340 criteria. It integrates two-thirds hard statistical data with one-third executive opinion survey data. The 2025 edition of the WCR particularly emphasizes that traditional determinants of competitiveness, such as macroeconomic stability, business-friendly environments, and quality infrastructure, are necessary but no longer sufficient. It highlights the increasing importance of digital readiness, green transition management, and sophisticated approaches to resilience for contemporary competitiveness, reflecting a growing recognition of future-oriented capacities. The WCR's findings underscore that strategic agility and wise policy are crucial for nations to "re-earn" competitiveness in a fragmented global economy (IMD (2025).)

The Global Innovation Index (GII) published by the World Intellectual Property Organization, assesses the innovation performance of approximately 130 economies, providing a comprehensive picture of innovation through around 80 indicators. These indicators cover the policy environment, education, infrastructure, and knowledge creation within each economy. Since its inception in 2007, the GII has significantly shaped the innovation measurement agenda, leading governments to systematically analyze their results and design policy responses to improve their innovation performance (WIPO (2025).)

While a competitive economy and innovative firms constitute essential components of future readiness, they represent only part of a broader and more complex picture. An economy may exhibit high levels of competitiveness under current conditions (characterized by today's technologies, geopolitical configurations, and environmental constraints) yet remain profoundly fragile and vulnerable to financial crises, environmental shocks, or disruptive technological change. For this reason, it is necessary to move beyond the conventional focus on competitiveness and innovation. To address this gap, we propose the development of a Future Readiness Index, designed to capture more comprehensively the capacity of economies to withstand, adapt to, and transform in response to structural disruptions.

The structure of the paper is as follows. Section 2 outlines the theoretical foundations of Future Readiness, introducing the key conceptual distinctions between **resilience** and **adaptive capacity**, and examining how these dimensions correspond to an economy's response to temporary versus permanent shocks. Section 3 highlights the differentiated roles of three core societal actors (governments, businesses, and citizens) whose collective readiness underpins national preparedness for future disruptions. Section 4 presents the six constituent elements that underly the Index of Future Readiness (IFR) that correspond to the resilience and adaptability of each societal actor. This section also explains how these elements can be systematically aggregated into sub-indexes to enable disaggregated analysis and policy relevance. Section 5 specifies the indicators associated with each of the six index elements, discussing the rationale for their selection, potential data sources, and methodological considerations. Section 6 discusses the structure and weights of the proposed index. Section 7 concludes.

## 2. Future Readiness: Beyond Resilience

The global financial crisis of 2008 and the subsequent Eurozone sovereign debt crisis (2010–2012) exposed significant structural fragilities within the European Union. These crises underscored the inadequacy of existing frameworks and compelled the Union to reconceptualize its approach to economic governance, social cohesion, and long-term strategic planning.

The crisis period revealed a series of systemic weaknesses that stressed the fragility of the European project. The Eurozone had no robust crisis-management instruments at its disposal: there was no banking union, fiscal coordination remained weak, and the EU budget offered only a limited capacity to respond. Member States such as Greece, Portugal, Ireland, Spain, and Italy slipped into deep and prolonged recessions, accompanied by soaring unemployment, demonstrating how a shock in one part of the Union could destabilize the monetary system as a whole. In this context, the long-standing reliance on stability as a guiding principle proved inadequate. Stability alone could not shield the Union from crisis. Shocks, it became clear, were inevitable, and the policy paradigm began to shift towards resilience: the capacity not only to withstand disruption, but also to recover in its aftermath.

Before the 2008 financial crisis, European Union policies were primarily designed to prevent crises through a framework of rules, most notably the Stability and Growth Pact and the Maastricht criteria. The experience of the crisis, however, demonstrated that rules alone were insufficient to shield the Union from systemic shocks. What was needed instead was the capacity to absorb disruption, and recover effectively. This realization brought the concept of resilience into the EU's policy vocabulary. In the social domain, resilience became associated with protecting households and workers from asymmetric shocks. Within cohesion policy, it translated into funding aimed at strengthening regions' ability to withstand future crises.

Between 2012 and 2014, the European Commission began to introduce the concept of resilience into its policy papers, using it initially in reference to crisis-hit regions and vulnerable social systems. This marked the first step in a gradual shift towards embedding resilience in the Union's policy framework. For example, in 2012 published the "Action Plan for Resilience in Crisis Prone Countries: 2013-2020" (European Comission (2012),) which reiterated the intent to build resilience in external interventions, focusing on helping vulnerable populations to withstand, cope, and recover from shocks effectively. In 2014, the Commission published the "EU Resilience Compendium" (European Comission (2014),) whose stated aim is "to build resilience to stresses and shocks," particularly in external, humanitarian contexts. In 2016, the *EU Global Strategy* elevated resilience to a core principle, applying it above all to external policy (particularly in relation to fragile states and neighborhood policy) before the concept gradually spread into internal governance.

The imperative of building resilient societies acquired heightened significance during the COVID-19 crisis. The pandemic exposed the inherent vulnerabilities of global value chains, demonstrating that economic interdependence could rapidly become a source of fragility. Lockdowns imposed in one part of the world precipitated cascading disruptions elsewhere, as critical parts and components remained immobilized in ports and on vessels stationed in countries under restrictive measures. This revealed the extent to which localized shocks could generate systemic consequences for the functioning of the global economy.

Immediately following the covid pandemic, in 2022, the Institute of Future-Fit Economies (or ZOE) published the "Framework for Economic Resilience" (see Hafele, Barth, Le Lannou, Bertram, Tripathi, Kaufmann, Engel, (2022).) Building on the definition of Vugrin, Warren, Ehlen, and Camphouse, (2010), Hafele et al (2022) define resilience as the "*magnitude and duration of the deviation from the targeted system* *performance level*" (page 18). In other words, resilient economies can be understood as those capable of returning to their intended trajectory following a shock within a relatively short period of time and with limited adverse effects on the various economic actors involved.

One year later, in 2023, Hafele, Bertram, Demitry, Le Lannou, Korinek, Barth, (2023) build on the ideas developed by Hafele et al (2022) to construct the "Economic Resilience Index" (ERI). The ERI's main goal was to assess the resilience of 25 European Member States during the two years of the covid-19 pandemic. The ERI was composed of 27 indicators divided into six "resilience dimensions": economic independence, education and skills, financial resilience, governance, production capacity, and social cohesion.

The word "resilience" has its roots in the Latin verb "resilire" which means "*to leap back, to spring back, to recoil*" ("re" means back or again and "salire" means to jump or leap). The first uses of the word resilience in English, dating back to the 1620s, referred to the physical properties of materials: the capacity of a spring, a piece of wood, or a metal to "spring back" after being bent or compressed. In the nineteenth century, engineers extended the term to describe a material's ability to absorb energy elastically before undergoing permanent deformation.

By the mid-nineteenth century, the concept had begun to migrate into medicine, where physicians used it to characterize the body's recovery from illness or injury. In the twentieth century, particularly between the 1950s and 1970s, psychologists adopted resilience to describe the capacity of children and individuals to adapt to and recover from trauma, stress, or adversity. Classic studies examined, for example, how some children were able to thrive despite growing up in poverty or under conditions of war.

The term underwent another significant transformation in the 1970s, when the Canadian ecologist C.S. Holling introduced it into environmental science (Holling (1973).) He defined resilience as the ability of an ecosystem to absorb disturbances such as fires, floods, or human interventions while still maintaining its core functions.

This marked a conceptual shift away from the idea of stability and towards adaptability as a measure of system health.

From the 1980s onward, resilience entered the vocabulary of the social sciences, particularly in the context of recurring crises such as the oil shocks of the 1970s, financial crashes, and climate-related disruptions. In economics, the term came to signify the capacity of economies to withstand temporary shocks, such as recessions. In parallel, the notion of community and societal resilience developed, denoting the ability of groups, cities, and nations to recover from large-scale disruptions such as earthquakes, wars, or pandemics.

In this paper, we advance a discussion on the concept of Future Readiness. While acknowledging that resilience constitutes a critical dimension of future readiness, we contend that resilience, understood in isolation, represents an incomplete framework. The optimal response of an economy to disturbances is contingent on the nature of the shocks it experiences. When confronted with temporary shocks, it is preferable for a society to resist, absorb the disturbance, minimize the resulting damage, and restore economic activity to its prior trajectory as swiftly as possible. In such cases, resilience (defined as the ability to "bounce back"), constitutes the appropriate response (Briguglio (2009), Hallegatte (2014), Brinkman et al. (2017), Capoani et al. (2025).).

Temporary shocks are commonly defined as short-lived disturbances that exert non-permanent effects on individual earnings or aggregate economic performance (Lucas (1977), Nelson and Plosser (1982), Barro (1984), Blanchard and Quah (1989), Bank for International Settlements, 2016). Their impact, while potentially severe, dissipates over time, allowing the economy to converge back to its pre-shock path or equilibrium (Kitsos, 2021). Importantly, the temporary character of these shocks should not be equated with triviality. As demonstrated by the Great Recession and the COVID-19 crisis, temporary shocks may generate extraordinarily large disruptions to economic systems. Additional examples include natural disasters such as hurricanes, earthquakes, crop failures, floods, wildfires, and tsunamis, as well as supply chain disruptions akin to those observed during the COVID-19 pandemic. Policy-induced disturbances can also function as temporary shocks, including tariff disputes and trade wars, sudden stops of capital flows, travel bans, trade embargoes, hikes in global interest rates, or even armed conflicts among key trading partners (Hardin, 2025; Freightfox, 2025).

However economies often face permanent shocks (Lucas (1977), Nelson and Plosser (1982), Barro (1984), Blanchard and Quah(1989).) That is, disturbances that permanently and fundamentally change the foundations of economies or even entire societies. The effects of these shocks are so profound that the optimal response to these types of shocks is not to resist, absorb and "bounce back" to the previous trend but to "bounce forward" by fundamentally transforming important elements of the economy.

The introduction of the automobile in the early twentieth century provides a vivid illustration of how technological innovations can fundamentally restructure entire economies. Prior to the spread of motor vehicles, a vast ecosystem of occupations revolved around the use of horses for transport and commerce. Farriers and blacksmiths shoed horses and repaired tack and iron fittings, while harness makers and other craftsmen specialized in the construction and maintenance of saddles, harnesses, and reins. The upkeep of horses in transit required stable ostlers, who fed, watered, and cared for animals at inns and staging posts. Passenger mobility depended on coachmen and carriage drivers, who operated both private coaches and public conveyances. The trade itself was sustained by horse dealers, who facilitated sales and auctions for transport purposes. In addition, a specialized manufacturing sector produced the material infrastructure for horse-drawn mobility: wagon makers and carriage builders constructed carts, stagecoaches, buggies, and hansom cabs, while wheelwrights designed and repaired wooden and iron-rimmed wheels. The reliance on animal transport also shaped urban services. Street cleaners, often referred to as "scavengers," were employed to remove horse carcasses from city streets, while manure collectors were responsible for handling animal waste, which was frequently resold as fertilizer. This extensive network of trades demonstrates the degree to which pre-automobile societies depended on equine-based transport systems.

The advent of the automobile had a profound and ultimately devastating impact on the extensive network of professions linked to horse-based transport, leading to the disappearance of virtually all of them. Had policymakers in 1905, confronted with rising unemployment in the equine sector, chosen to prioritize resilience by shielding their societies from the diffusion of the automobile in the expectation that the economy would eventually "bounce back" to its previous state, such a strategy would have been gravely misguided. The optimal course of action would have been to prepare citizens, businesses, and governments for the new opportunities generated by the automobile economy. For businesses, this would have implied readiness not only to invest in the direct manufacturing of cars but also to engage in complementary industries, including oil and gas for internal combustion engines and the rubber industry for tire production. For governments, the priority should have been the planning and construction of new infrastructure (such as paved roads, highways, and city streets) capable of supporting motorized transport. For citizens, adaptation would have required a willingness and capacity to transition from traditional horse-related occupations to emerging roles in the automobile sector, ranging from factory work to engineering and design.

This historical episode of technological transformation demonstrates that future readiness requires more than resilience understood in the narrow sense of absorbing shocks and restoring a prior equilibrium. Instead, it demands that societies develop the capacity to adapt and to reconfigure their economic structures when confronted with disruptions that fundamentally and permanently alter the foundations of economic life. While countries that are "future ready" must indeed maintain mechanisms to mitigate the adverse consequences of technological change (for example, social protection systems designed to address "technological unemployment" (Keynes (1930),) coping strategies alone are insufficient for preparing societies to confront the future.

To be truly future ready, a country must also possess the ability to reallocate resources (capital, labor, institutions, and technology) from sectors destined to disappear to those emerging as a result of new technological paradigms. In this sense, future readiness can be understood as a function of economic flexibility: the capacity to facilitate resource transfers and to adapt effectively to new environments. We refer to this dimension as **adaptive capacity[1]**, which, in conjunction with resilience, constitutes a necessary component of the proposed Future Readiness Index.

Technological progress is not the only driver of permanent transformations that necessitate the restructuring of economic systems. Other enduring disturbances exert similarly profound effects. Climate change, for instance, represents a structural challenge with wide-ranging implications for production, consumption, and long-term sustainability (Smith, 2025; Tol, 2018). Demographic shifts likewise play a critical role: declining fertility rates combined with rising life expectancy are reshaping global population structures, resulting in shrinking working-age cohorts and rising dependency ratios. These demographic dynamics place downward pressure on long-term economic growth, strain public pension systems, and require fundamental reconfigurations of labor markets and patterns of consumption (Reggiani et al., 2001). In addition, geopolitical realignments (manifested in persistent trade tensions, economic sanctions, or significant shifts in global power relations) can generate prolonged disruptions in supply chains, elevate operational costs, and alter the structure of international trade.

---

[1] The concept of antifragility, introduced by Nassim Taleb broadens the conventional understanding of adaptive capacity by extending it to temporary shocks as well. Unlike resilient systems that merely recover from disruption, antifragile economies improve when exposed to stressors, using volatility as a source of renewal and growth. This perspective highlights a crucial insight: a society's adaptive capacity must be designed not only to endure both temporary and permanent shocks, but to learn from them and emerge stronger. In this light, adaptive capacity is not simply a defensive attribute. It is a strategic capability central to long-term dynamism and national preparedness (Taleb (2012).)

Such dynamics carry lasting implications for national economies, underscoring that resilience alone is insufficient: societies must also be capable of systemic adaptation and transformation in the face of permanent structural change.

It is important to emphasize that the capacity to adapt and transform an economic system should not be conceived solely in reactive terms, as a response implemented only after a permanent shock has already materialized. Rather, it must also be understood as a proactive capability, encompassing both the institutional mechanisms and the strategic vision necessary to anticipate future developments and to contribute actively to their realization. In the context of technological progress, for example, a future-ready country should not limit itself to passively adopting innovations introduced elsewhere and adapting to a future shaped by others. Instead, it should possess the capacity to generate innovation endogenously and to play an active role in shaping the technological trajectory, thereby constructing its own future rather than merely adjusting to external transformations.

## 3. Three Key Actors of Future Readiness: the GBC model

It is tempting to think that governments should be responsible for a country's future-readiness. Indeed, governments everywhere are trying to keep up with technological developments and protect their people from an increasingly unpredictable world. But how can they create the agility and resilience required to be ready for the future?

In his book *Government Reimagined: Leading Through New Realities*, Ali Qassim Jawad (2020), governments do not work in isolation (or exist simply to keep civil servants in gainful employment). Governments are part of an ecosystem that includes business and citizens. He refers to this as the GBC (Government-Business-Citizen) model. Governments on their own cannot legislate for future readiness. Nor can either of the other two parts of the ecosystem in isolation create the pre-conditions that enable a country to respond to disruptive change.

In short, the relationship between government, business and the citizenry is symbiotic; together the three component parts form a network of relationships that binds the culture of a healthy nation, and also determines the degree of preparedness for unexpected disruptions. This became powerfully clear during the Covid-19 pandemic. The pandemic was a stark reminder of how quickly the world can change, and events can escalate. Many governments were caught unprepared. Yet despite the mistakes that were made, there were examples where business and citizens showed their willingness to play their part, whether it was firms reconfiguring production lines to supply personal protective equipment, or citizens voluntarily self-isolating to protect the vulnerable members of their communities.

What this demonstrates is the adaptability and ingenuity of human beings and the interconnectedness of the GBC model. Government alone cannot hope to have all the solutions or even to know all the questions that would allow it to anticipate and respond to future disruptions. What is needed to create a thriving nation is intelligent collaboration between the public and private sectors, which also recognises the public duty of private citizens. When all the actors are collaborating effectively, we will have the conditions for a progressive and thriving country that is also future ready.

The government is sometimes blamed for a nation's ills, but the reality is both more nuanced and more exciting. Viewed through the lens of the GBC model, the responsiveness and responsibility for a nation's future preparedness rests with all three actors. This realization opens the door to new opportunities. It also provides a compelling framework for a future readiness index.

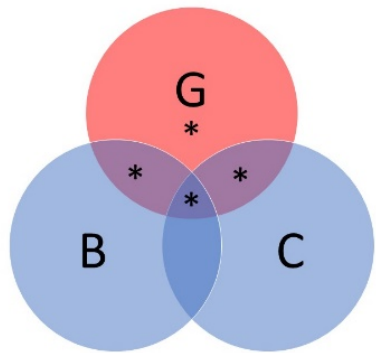

The two-way interactions between government and business and government and citizens are represented by the overlap between the G and B circles and the G and C circles, respectively.  But there is also growing scope for three-way collaboration, which is denoted by the asterisk at the centre of the model where all three circles overlap.

Let us drill down deeper into government's interactions with business and citizens. As stated before, this is a symbiotic ecosystem, where all three parts rely on each other for their success. For example, businesses need citizens as employees and customers, and need government to create the legal framework, infrastructure and conditions for trade. Citizens, on the other hand, require meaningful work and goods and services from business and the personal security and services that the state provides. And government relies on the talents of its citizens as civil servants and technocrats, and on an energetic and profitable private sector to help fund, innovate and support public services.

When the ecosystem works effectively, it's a win-win: win scenario. It is therefore essential that the three component parts of the nation can collaborate in productive and purposeful ways. Collaboration, however, relies on trust. And building trust between government, business and citizens takes time and hard work. It is far easier to create rules and red tape. But excessive regulation does not engender future readiness. A rigid system cannot respond rapidly to new challenges and opportunities. Loss of trust in government on the part of business and citizens manifests itself not only in a growing cynicism towards politicians and government leaders but also in a diminishing trust in the institutions of state. To create a thriving GBC ecosystem, it is essential to reclaim and rebuild that institutional trust.

The nations that will flourish in the decades ahead will be those that create a harmonious balance between government (which protects and enables), business (the wealth creation engine) and the citizen (who contributes their skills and loyalty). These three parts of society are in a symbiotic relationship and must function collaboratively. To do so, however, government leaders will have to find ways to foster trust and cooperation between all three. The inter-relationship between these three actors is central to a nation's future readiness.

The **government and public sector** are critical in providing strategic leadership and establishing the enabling environment for Future Readiness. Governments play a crucial role in establishing the macroeconomic and institutional buffers necessary for resilience to temporary shocks by providing the necessary fiscal buffers, the monetary space, the emergency response systems, and the necessary physical and digital infrastructures. Governments are also central to creating an enabling environment for long-term adaptation by providing institutional flexibility, innovation foresight, green infrastructure, and flexible labor rules that allow the transfer of human capital from old to new economic sectors.

The **business and private sector** contributes significantly to resilience through the provision of capital buffers, flexible supply chains, business planning, and sectoral diversification. It also contributes to the adaptive capacity of a nation through research and development and other forms of innovation, market dynamism, and strategic adaptation.

Finally, the **citizenry**, encompassing individuals and civil society organizations (CSOs), must show resilience through social safety nets, emergency savings and financial buffers, and financial and risk management literacy. The citizenry must show its capacity to adapt to long-term shocks through live-long learning and skill development (post education re-skilling and up-skilling), its sectorial and geographic mobility, its proactive financial planning and its technological literacy and connectivity.

The effectiveness of a nation's Future Readiness hinges on the dynamic interaction and collaboration among these three entities. Multi-stakeholder participation, trust-building, effective communication, and information sharing provide the enabling conditions to enhance national resilience. Governments, businesses, and communities must integrate resilience and adaptive capacity into decision-making at all levels, fostering an environment where citizens have a strong sense of trust in public systems and institutions.[33] This integrated approach ensures that resources, knowledge, and capabilities are mobilized across society, transforming individual and sectoral strengths into a cohesive national capacity for future success.

Achieving and sustaining national Future Readiness is not the sole responsibility of any single entity but demands a unified, "whole-of-society" approach (Jawad 2020).) This comprehensive endeavor requires the synergistic engagement of the government (and the public sector generally), businesses (including the private sector), and the citizenry. Each plays a distinct yet interconnected role in building a nation's capacity to confront temporary shocks and adapt to permanent changes.

# 4. The Six Elements of the FRI

In Section 2, we outlined the necessity of incorporating both resilience and adaptive capacity into a comprehensive index of future readiness. Section 3 further emphasized the central role of the three principal actors within an economic system: government, the business sector, and the citizenry. Taken together, these considerations imply that an index of future readiness should be structured around six components, corresponding to the intersection of these two dimensions with the three actors. Specifically, the index should include: Government Resilience (GR), Business Resilience (BR), Citizens' Resilience (CR), Government Adaptive Capacity (GA), Business Adaptive Capacity (BA), and Citizens' Adaptive Capacity (CA). A summary of these six elements is presented in Table 1.

**Table 1: The Six Elements of the Future Readiness Index**

| | Resilience (Short-Term Shocks) | Adaptive Capacity (Long-Term Shocks) |
|---|---|---|
| **Government** | GR | GA |
| **Business** | BR | BA |
| **Citizens** | CR | CA |

## 4.1. Government's Role in Building Resilience (GR)

Governments play a crucial role in establishing the macroeconomic and institutional buffers necessary for resilience to temporary shocks (Rose (2004), (2007), 2009)). There are five areas in which the government should contribute to the resilience of a national economy.

**Fiscal Buffers:** Maintaining prudent fiscal policies, including manageable debt levels, balanced budgets, diversification of the tax base and robust domestic revenue mobilization, creates fiscal space for rapid responses to disasters and economic downturns. This allows governments to fund emergency relief and support recovery efforts without jeopardizing long-term economic stability (Rezzai et al (2025).)

**Monetary Policy Space:** Central banks contribute to resilience by building credibility in the monetary system to counter demand shocks and by maintaining the capacity to implement expansionary monetary policy without creating excessive inflation or international payment disequilibria (Rezzai et al (2025).)

**Emergency Response Systems:** Investing in early-warning systems, disaster preparedness, and robust public infrastructure helps limit immediate production losses and facilitates rapid reconstruction and recovery after events like natural disasters (Cutter et al (2008), Capoani et al (2025).)

**Infrastructure Resilience (including Digital infrastructure):** Governments have a primary responsibility to ensure that critical infrastructures are both robustly designed and adequately maintained. This entails incorporating redundancy into transport networks, energy grids, and water systems in order to prevent localized failures from cascading into systemic breakdowns. Preventive maintenance, together with systematic upgrading, is essential to mitigate the vulnerabilities associated with aging assets.

In addition to maintenance, the state must prioritize investment in modernization, embedding resilience standards within the planning and execution of new infrastructure projects. Increasingly, this requires climate-proofing assets against risks such as floods, heat waves, and rising sea levels. The deployment of smart technologies and digital monitoring systems further enhances resilience by enabling the anticipation of risks and the capacity to respond to shocks in real time.

Digital resilience requires that governments safeguard and strengthen core infrastructures such as broadband, 5G, and satellite systems. Critical sectors (including finance, health, energy, and transport) depend on their reliability. Ensuring redundancy, interoperability, and minimum cybersecurity standards for both public and private operators is therefore essential to guarantee continuity of services during disruptions.

Equally important is the development of strong cybersecurity and threat management capacities. Governments must establish national strategies, specialized agencies, and rapid-response teams while also cooperating internationally, since most cyber risks transcend borders. Anticipating new vulnerabilities created by technologies such as artificial intelligence or quantum computing is crucial to maintaining security and reducing systemic risks.

**Trade and Currency Buffers:** In the context of economic resilience, trade and currency buffers enable countries to withstand shocks such as capital outflows, exchange rate volatility, and trade disruptions. Governments play a central role in building these buffers through prudent macroeconomic management. Maintaining adequate foreign exchange reserves and establishing currency swap lines with major central banks provide liquidity and stabilize markets in times of stress. Trade diversification (both in products and geographic partners) reduces vulnerability to sector-specific downturns or geopolitical disruptions. Flexible exchange rate regimes can further absorb external shocks, provided they are supported by credible fiscal and monetary policies. Prudent external debt management, particularly limiting reliance on short-term foreign-currency borrowing, and the development of domestic capital markets strengthen resilience against sudden stops. Ultimately, institutional credibility (anchored in transparent monetary policy, consistent fiscal frameworks, and clear communication) remains essential to sustaining investor confidence and mitigating systemic risk.

## 4.2 Businesses' Role in Building Resilience (BR)

Businesses must implement strategies to withstand and recover from short-term disruptions:

**Robust Risk Management and Capital Buffers:** Financial institutions and firms need strong risk management frameworks and sufficient capital and liquidity buffers to absorb economic and financial shocks. Well-capitalized banks, for instance, are more resilient and less likely to amplify negative macroeconomic shocks (Mester (2024).).

**Supply Chain Redundancy and Flexibility:** Building resilient supply chains involves rapid detection and response, end-to-end data-driven control, and strategic redundancies such as emergency stockpiles, safety stocks, and diversified sourcing from multiple suppliers. Agile manufacturing processes and adaptable logistics networks are crucial for quickly shifting production or transportation routes

**Business Continuity Planning:** Ensuring businesses understand their vulnerabilities and are prepared to resume operations swiftly after an event is critical for minimizing downtime and losses (U.S. Development Administration, (2025).)

**GDP Diversification**: in the context of economic resilience, GDP diversification denotes reducing reliance on a narrow set of sectors, products, or markets so that shocks in one area do not escalate into systemic crises. Governments contribute by fostering an enabling environment through investment in education, research, and infrastructure, by implementing industrial and trade policies that encourage value-added production and market diversification, and by deploying fiscal and financial tools to support emerging sectors such as green and digital industries. Businesses, in turn, strengthen diversification resilience by expanding product portfolios, broadening supply chains, investing in innovation and technology, and pursuing internationalization strategies. Ultimately, diversification rests on the interaction between state and market actors: governments establish the framework conditions, while firms provide innovation and adaptive capacity. Together, these efforts reduce structural vulnerabilities and position economies to absorb shocks.

## 4.3. Citizens' Role in Building Resilience (CR)

Individuals and households contribute to overall economic resilience by enhancing their capacity to absorb and recover from personal economic shocks:

**Social Safety Nets:** Robust social protection systems, including unemployment benefits, health coverage, and welfare programs, provide crucial support to households during temporary income losses or health emergencies, enabling them to maintain stability (Hallegatte (2014).)

**Emergency Savings and Financial Buffers:** Maintaining adequate emergency cash savings (e.g., three to six months of income) provides a financial cushion during difficult times, reducing reliance on external support and enabling individuals to smooth consumption (Hallegatte (2014).)

**Financial Literacy and Risk Management:** Understanding personal finances, managing debt, and utilizing insurance mechanisms (e.g., health, unemployment) help individuals smooth shocks over time and mitigate the impact of unexpected events.

**Social Capital and Trust.** In economic and social systems, social capital and trust are fundamental to national resilience, as they facilitate cooperation among individuals, communities, and institutions. Trust among citizens fosters collective action in times of crisis, reducing coordination costs and enabling rapid responses to disruptions. Equally, trust in institutions enhances the legitimacy and effectiveness of policy interventions by encouraging voluntary compliance, whereas weak trust undermines resilience by fostering resistance and social fragmentation.

## 4.4. Government's Role in Fostering Adaptive Capacity (GA)

As mentioned above, a permanent shock is fundamentally defined as a long-lasting change within an economy that alters its output potential or long-term growth trajectory. Unlike temporary shocks, which may cause short-term fluctuations but do not affect long-term capacity, permanent shocks have enduring effects on critical macroeconomic factors such as productivity, labor supply, and technology, leading to a sustained shift in the aggregate supply curve. Such shocks impact not only immediate economic performance and cash flows but also fundamentally reshape future prospects.

The distinction between permanent and temporary shocks is crucial for understanding economic responses. Temporary shocks are short-lived, with their effects dissipating over time, allowing the economy to eventually return to its pre-shock path. In contrast, permanent shocks induce a "new equilibrium state" that is enduring, causing irreversible changes to the economy's growth ceiling. Empirical evidence consistently demonstrates that economic shocks can permanently lower employment in affected regions, with output losses often not regained on average, contributing to lower long-run growth. This enduring scarring effect of permanent shocks highlights the imperative for **structural adaptation**. Many significant shocks are not temporary deviations but rather permanent structural shifts. This means that traditional "**bounce back**" resilience is insufficient. Instead, economies must possess **adaptive capacity** to undergo fundamental structural transformation, or "**bounce forward**," to navigate these new, often lower, equilibrium paths or proactively shift to more favorable ones. This elevates the importance of policies that facilitate structural change rather than just short-term stabilization (Brinkmann et al (2017).)

Governments are central to creating an enabling environment for long-term adaptation:

**Institutional Flexibility and Adaptive Governance:** Fostering high-quality, flexible institutions and open, collaborative governance processes enables timely policy reforms and the mobilization of public resources in response to long-term shifts.[1] This includes strategic foresight capabilities to anticipate future challenges (U.S. Development Administration, (2025).)

**Investment in Innovation and Green Infrastructure:** Directing public investment towards research and development (R&D), emerging technologies, and climate-smart infrastructure (e.g., renewable energy, resilient urban planning) facilitates structural transformation and addresses long-term environmental and technological changes (Capoani (2025).)

**Labor Market Flexibility:** Implementing policies that promote labor market flexibility (ease of hiring/firing, retraining systems, labor mobility) allows the workforce to adjust to changing demands and transition into emerging sectors. This must be balanced with worker protections to ensure social stability (OECD (2008).)

## 4.5. Businesses' Role in Fostering Adaptive Capacity (BA)

Businesses must proactively embrace innovation and strategic shifts to adapt to permanent changes:

**Innovation Systems and R&D:** Investing in R&D, fostering an entrepreneurial culture, and developing new products and services are crucial for adapting to technological revolutions and shifting consumer preferences. This includes developing adaptive marketing capabilities to respond swiftly to evolving customer needs (Barro and Sala-i-Martin (2004).)

**Market Dynamism and Diversification:** Cultivating dynamic markets that can rapidly reallocate resources across sectors and diversifying the economic base reduces reliance on single industries, making economies more resilient to structural shifts. This also involves adapting supply chains through strategic diversification of sourcing (Brinkmann (2017).)

**Strategic Adaptation to Economic Cycles:** Businesses need to tailor strategies according to prevailing economic conditions, focusing on growth during expansion and cost optimization during downturns, and continuously monitoring performance.

**Capital Flexibility.** The ability of firms to reallocate capital from declining to emerging sectors is a key determinant of an economy's future readiness. By preventing resources from remaining locked in low-growth activities and channeling them toward innovation-driven industries, firms facilitate structural transformation, sustain employment, and generate new sources of growth. This capacity enhances adaptability to permanent shocks and underpins long-term future readiness.

## 4.6. Citizens' Role in Fostering Adaptive Capacity (CA)

Individuals must cultivate personal adaptive capacities to thrive in a changing world:

**Lifelong Learning and Skill Development:** Engaging in continuous education, re-skilling, and up-skilling programs is fundamental for individuals to acquire new competencies as labor markets evolve due to technological advancements or demographic shifts. A highly skilled and adaptable population is a resilient population (Brinkmann (2017).)

**Connectivity:** the access to networks, information, and social relations constitutes a critical determinant of the citizenry's capacity to adapt to new sectors and changing circumstances. Greater connectivity enhances the diffusion of knowledge and skills, allowing individuals to recognize emerging opportunities, acquire new competencies, and transition into alternative forms of employment. Digital connectivity, in particular, provides access to online education, labor market platforms, and professional networks, thereby facilitating re-skilling and improving job matching.

**Labor Mobility:** The willingness and ability to transition between sectors or relocate geographically in response to shifts in job availability are crucial for adapting to structural changes in the economy (Zehra (2025).)

**Proactive Financial Planning:** Beyond emergency savings, individuals should engage in long-term financial planning that accounts for potential permanent shifts in income, retirement needs, and consumption patterns, adapting investment strategies accordingly (U.S. Bank (2025).)

## 4.7. Actor-Specific and Theme-Specific Readiness Sub-indexes

An interesting feature of our approach is that the six elements identified in this section can be restructured to produce three actor-specific sub-indexes, and two theme-specific sub-indexes. These sub-indexes enable a more granular or disaggregated assessment of future readiness and are summarized in Table 2.

**Table 2: Combining the Six Elements of the IFR to construct interesting sub-indexes**

| | Resilience (Short-Term Shocks) | Adaptive Capacity (Long-Term Shocks) | SUM (Actor-Specific Readiness) |
|---|---|---|---|
| **Government** | GR | GA | GR + GA = Government Future Readiness |
| **Business** | BR | BA | BR + BA = Business Future Readiness |
| **Citizens** | CR | CA | CR + CA = Citizens Future Readiness |
| **SUM (Thematic)** | GR + BR + CR = National Resilience | GA + BA + CA = National Adaptive Capacity | GR+BR+CR+GA+BA+CA = National Future Readiness |

A **Government Future Readiness** can be calculated by averaging Government Resilience (GR) and Government Adaptability (GA).

A **Business Future Readiness** can be derived by averaging Business Resilience (BR) and Business Adaptability (BA).

A **Citizen Future Readiness** can be calculated as the average of Citizen Resilience (CR) and Citizen Adaptability (CA).

This aggregation method enables a clear and comparable view of the overall preparedness of each actor to withstand temporary disruptions and adapt to permanent changes. It also facilitates targeted analysis and policy interventions by identifying sectoral strengths and vulnerabilities within the broader Future Readiness framework.

In addition to sectoral analysis, the six elements of the Future Readiness Model can also be aggregated vertically to produce two thematic sub-indexes:

**National Resilience Index (NRI):** This index captures the collective capacity of all three societal actors (government, business, and citizens) to withstand temporary shocks. It is calculated as the average of Government Resilience (GR), Business Resilience (BR), and Citizen Resilience (CR).

**National Adaptive Capacity Index (NAI):** This index reflects the collective ability of the same actors to adapt to long-term, structural transformations and permanent shocks. It is calculated as the average of Government Adaptability (GA), Business Adaptability (BA), and Citizen Adaptability (CA). These two sub-indexes (NRI and NAI) offer a complementary perspective by emphasizing the nature of the challenges faced (short-term vs. long-term), rather than the sector of society involved.

**Final Aggregation: National Future Readiness Index (NFRI)** The overall National Future Readiness Index (NFRI) can be calculated using either of two equivalent aggregation methods:

The sum (or average) of the National Resilience Index (NRI) and the National Adaptive Capacity Index (NACI).

The sum (or average) of the Government Future Readiness, Business Future Readiness, and Citizen Future Readiness sub-indexes.

In both cases, the resulting index reflects the same total contribution of all six underlying elements:

NFRI=GR+BR+CR+GA+BA+CA

This dual approach allows policymakers and analysts to view future readiness either through the lens of sectoral responsibility or shock response type, providing flexibility for both strategic planning and targeted intervention.

## 5. Indicator Framework: Proposed Metrics for Each Element

This section outlines a preliminary set of proposed indicators, identified as critical metrics for assessing the capacity of each key actor (government, business, and citizens) to either resist temporary shocks or adapt to permanent structural changes. This compilation represents a set of preliminary metrics for inclusion in a robust future readiness index, developed independently of current data availability constraints. While some measures may not yet be universally accessible, it is anticipated that many exist in some form within current national and international data repositories. For context, factors often included in existing indices are conceptually acknowledged, though their specific categorization may differ in this proposed framework.

### GR: Government Resilience

Remember that this element captures the readiness of government institutions and the broader public sector to withstand temporary shocks, ensuring short term flexibility and preventing systemic collapse. GR reflects the strength of emergency response systems, continuity of governance mechanisms, fiscal flexibility, and institutional stability during periods of acute stress.

**GR1: Fiscal Buffers and Monetary Policy Space**

Inflation Rate

Public Debt to GDP Ratio

Budget Balance as a Fraction of GDP

Sovereign Credit Rating

**GR2: Diversification of Government Revenue**

Percentage of Government Revenue from a Single Dominant Source (e.g., oil)

Variability of Government Revenue over the last five years

**GR3: Infrastructure and Digital Resilience**

Reliability of energy, transport, and communication systems

Cybersecurity of government and public digital infrastructure

Sovereign Wealth Fund capacity (e.g., as a percentage of GDP, indicating ability to finance emergency rebuilding)

**GR4: Diplomatic Diversification**

Measure of multilateral engagement (e.g., number of international agreements, participation in international organizations)

Visa-free travel access for citizens (as a proxy for diplomatic reach and soft power)

Number of bilateral trade agreements in force

**GR5: Trade and Currency Buffers**

Foreign Reserves as a percentage of GDP (Reserve Adequacy)

Current Account Balance as a percentage of GDP

Exchange Rate Regime Flexibility

Trade as a fraction of GDP

## GA: Government Adaptive Capacity

This element captures the readiness of government institutions and the broader public sector to adapt when confronted with permanent shocks. GA reflects the ability to enact forward-looking policies, reform outdated systems, and mobilize public resources and support in response to long-term shifts that redefine the operating environment.

**GA1: Institutional Flexibility**

E-government Development Index (EGDI)

Voice and Accountability Index (World Bank Governance Indicators)

Strategic Foresight Capacity (e.g., existence and effectiveness of national foresight units)

Political Stability and Absence of Violence Index (World Bank Governance Indicators)

Rule of Law Index (World Bank Governance Indicators)

Open Data Implementation Index

Legal Frameworks (e.g., World Bank Doing Business indicators related to legal enforceability)

Responsive Policy Making (e.g., Polity Score, indicating democratic and adaptable governance)

Decentralization Index (e.g., fiscal and administrative decentralization)

Predictability of Regulatory Environment

**GA2: Adaptive Governance**

Regulatory Quality Index (World Bank Governance Indicators)

Government Effectiveness Index (World Bank Governance Indicators)

Perceived Burden of Bureaucracy (e.g., survey data)

**GA3: Labor Market Regulatory Flexibility**

Ease of Hiring/Firing (e.g., OECD Employment Protection Legislation Index)

Effectiveness of National Retraining Systems

Labor Mobility Index (e.g., inter-sectoral and geographical mobility facilitation)

**GA4: Environmental and Sustainability Governance (Investment in Innovation and Green Structure)**

Renewable Energy Share in Total Energy Consumption

Renewable Energy Adoption Rates

Climate Resilience Policies and Investments

Environmental Performance Index (EPI)

**GA5: Capacity for Policy Experimentation and Entrepreneurial State (this is related to the paper "Finding the Omani Way")**

Existence and utilization of regulatory sandboxes or pilot programs for policy innovation (e.g., mirroring Singapore or China's approach to testing prototype policies before nationwide implementation).

# BR: Business Resilience

This element captures the readiness of domestic businesses and markets to withstand temporary shocks.

**BR1: Robust Risk Management and Capital Buffers**

Banking System Solvency (e.g., Bank Capital Ratios)

Banking System Regulation and Oversight (e.g., Non-performing Loan Ratios, Financial Supervision scores)

Stock Market Capitalization as a percentage of GDP

**BR2: Infrastructure and Digital Resilience (Private Sector)**

Cybersecurity of private business institutions and critical infrastructure

**BR3: GDP and Trade Diversification**

Sectorial Concentration in GDP

Sectorial Concentration in Exports

**BR4: Supply Chain Redundancy and Flexibility**

Decentralization of energy systems

Decentralization of food production and distribution

Supply Chain Concentration Index (lower concentration indicates higher resilience)

Logistics Performance Index (LPI)

**BR5: Formality of Economic Activity**

Informal Economy Size (as a percentage of GDP, inversely related to resilience)

Black Market Premium (as an indicator of economic distortion)

## BA: Business Adaptive Capacity

This element assesses the private sector's capacity to adapt effectively to permanent shocks. Key factors include innovation, dynamism, and flexibility, which enable the efficient reallocation of economic, financial, and human resources from sectors adversely affected to those poised for growth in response to structural change.

**BA1: Innovation Systems and R&D**

Research & Development (R&D) Spending as a percentage of GDP

Educational Attainment (e.g., PISA scores, tertiary education enrollment)

Digital Infrastructure Access for Firms (e.g., internet access, broadband penetration)

Global Innovation Index (GII) ranking

Percentage of Firms with an Online Presence (Website)

**BA2: Market Dynamism**

SME Ecosystem Health (e.g., new firm entry rate, survival rate of startups)

Competition Policy Effectiveness (e.g., market concentration indices)

Entrepreneurial Culture Index (e.g., survey-based measures of entrepreneurial attitudes)

Ease of Closing a Business (World Bank Doing Business indicator)

Customs Clearance Time

Global Mobility Access (e.g., ease of international business travel)

Logistics Competence (e.g., LPI sub-components)

**BA3: Economic Complexity and Sophistication**

Economic Complexity Index (Hausmann)

High-Technology Exports as a percentage of total exports

**BA4: Capital Flexibility**

Venture Capital Investment as a percentage of GDP

Foreign Direct Investment (FDI) as a percentage of GDP

Domestic Credit to Private Sector as a percentage of GDP

**BA6: Institutions Supporting Business Experimentation and Learning**

Measures of regulatory sandboxes for business innovation, industry-academia collaboration indices.

## CR: Citizen Resilience

This element captures the readiness of families and individual citizens to withstand temporary shocks without suffering social collapse due to mass poverty or widespread unemployment.

**CR1: Social Safety Nets and Financial Security**

Social Spending as a percentage of GDP (Welfare Programs)

Unemployment Benefits Coverage and Adequacy

Health Coverage (e.g., percentage of population with access to healthcare)

Poverty Gap Coverage (e.g., percentage of the poverty gap covered by transfers)

Household Savings Rate

Cumulated Savings per capita

Average Percentage of Disposable Income Available after Housing Costs (rent or mortgage)

Progressivity of the Tax System

Account Ownership at Financial Institutions (percentage of population over 15)

**CR2: Emergency Savings and Financial Buffers**

Individuals savings rate

**CR3: Financial Literacy and Risk Management**

Percentage of population with financial literacy

**CR4: Social Capital and Trust**

Public Trust in Institutions (e.g., World Values Survey data)

Social Cohesion Index (e.g., World Values Survey data, measures of civic participation)

Gini Coefficient (as an inverse measure of income equality and potential social friction)

Civic Engagement Rates (e.g., volunteerism, participation in community organizations)

## CA: Citizen Adaptability

This element evaluates citizens' capacity to adapt to permanent shocks. Critical factors include the ability to acquire new skills as previous competencies become obsolete, as well as the flexibility to transition between sectors or relocate geographically in response to shifts in job availability caused by structural changes within the economy.

**CA1: Lifelong Learning and Skill Development**

Overall Educational Attainment (e.g., average years of schooling, tertiary enrollment rates)

Availability and Participation in Post-Education Re-skilling Programs

Availability and Participation in Post-Education Up-skilling Programs

Emphasis on Forward-Looking Education (e.g., curricula fostering curiosity, critical thinking, and creativity)

STEM Graduates per capita

Net Flow of International Students (indicating openness to global knowledge exchange)

Household Internet Penetration Rate

**CA2: Labor Mobility**

Sectoral Labor Mobility Rates

Geographical Labor Mobility Rates

**CA3: Population Dynamics**

Working Age Population Growth Rate

Urban Population Growth Rate (as an indicator of demographic shifts and potential for agglomeration effects)

**CA4: Digital Connectivity and Skills**

Digital Skills among the Population (e.g., digital literacy rates)

E-participation Index (e.g., citizen engagement in online governance)

Mobile Broadband 5G Subscriptions (households)

Internet Bandwidth Speed (households)

Access to Digital Payments (e.g., percentage of population using mobile money or online banking)

**CA5: Proactive Financial Planning**

Ability to engage in long-term financial planning

## 6. Weighting, Normalization Methods, and Averaging Schemes

For the aggregation of indicators within the proposed framework, we adopt weighting schemes, normalization procedures, and rescaling approaches consistent with established international practice, in line with OECD guidelines (OECD, 2008) and methodologies employed by leading indices such as the Global Competitiveness Index of the World Economic Forum, the Global Innovation Index of WIPO, and the World Competitiveness Index of the Institute for Management Development. Specifically, the framework applies equal weighting at each level of aggregation: individual measures within sub-elements, sub-elements within elements, and elements within their respective sub-indexes and the overall composite index.

Normalization and rescaling of all indicators to a 1–100 scale are introduced to ensure comparability and interpretability across heterogeneous datasets. For negatively oriented indicators, such as inflation or fiscal deficits, a min–max linear rescaling with inverted orientation is employed. This procedure linearly maps the worst observed value to the lowest score (1) and the best observed value to the highest score (100), thereby aligning directionality across all measures. Formally, the normalization is expressed as:

$$score_i = 1 + \frac{Max(X) - x_i}{Max(X) - Min(X)} \times (100 - 1)$$

For positively oriented indicators (where higher values are better like, for example, patents or R&D), we use a regular min-max indicator where, again, 1 is the worst possible score and 100, the best:

$$score_i = 1 + \frac{x_i - Min(X)}{Max(X) - Min(X)} \times (100 - 1)$$

## 7. Conclusion

This paper has proposed and elaborated a new framework for assessing a country's preparedness to navigate a future defined by uncertainty, complexity, and accelerating change. At its core, the Index of Future Readiness (IFR) offers a multidimensional measure of a nation's foundational strength. That is, its capacity not only to recover from shocks but also to adapt proactively and transform in response to long-term disruptions. As the 21st century continues to unfold with intensifying environmental, geopolitical, technological, and societal pressures, such a tool is urgently needed to complement (and in many respects, transcend) existing benchmarking instruments that remain largely backward-looking or limited in scope.

The IFR is built upon two foundational concepts: economic resilience and adaptive capacity. Resilience captures a system's ability to withstand and recover, or "bounce back" from temporary shocks such as natural disasters or financial volatility; adaptive capacity, by contrast, refers to the deeper and more structural ability to adjust, reorganize, and "bounce forward" in the face of enduring transformations such as demographic transitions, technological revolutions, or climate change. These two capacities are not only conceptually distinct but also synergistic: a nation that is merely resilient may return to a suboptimal path, while one that is purely adaptive without buffer mechanisms may suffer destabilization under acute pressure. Future readiness requires the integration of both capabilities.

Crucially, this readiness must be understood as a "whole-of-society" endeavor, involving the coordinated and complementary contributions of government, business, and citizens. The framework proposed here, structured as a 3×2 matrix of six core components, offers a novel way to disaggregate and assess these actors' capacities to respond to both temporary and permanent shocks. It recognizes, for example, that governments are not only first responders during crises but also system architects responsible for long-term policy foresight and institutional flexibility; that businesses contribute not just through healthy balance sheets and robust supply chains, but also through innovation ecosystems and market dynamism; and that citizens play an equally vital role by maintaining financial buffers, cultivating digital and cognitive skills, and remaining mobile and connected in changing labor markets.

This conceptual structure is operationalized into a composite index, allowing for the aggregation of granular indicators into actor-specific and theme-specific sub-indexes. In doing so, the IFR enables both vertical and horizontal analysis, providing a basis for targeted policy interventions (e.g., strengthening citizen adaptive capacity) as well as cross-national benchmarking of overarching strengths and weaknesses (e.g., comparing national resilience levels across countries). Its emphasis on disaggregated and future-facing capabilities provides a marked improvement over traditional competitiveness indexes, which often emphasize current performance and fail to incorporate the forward-looking, anticipatory elements necessary in today's environment.

Importantly, the IFR does not seek to replace existing indexes such as the Global Competitiveness Index, the World Competitiveness Rankings, or the Global Innovation Index. Rather, it builds upon their strengths while addressing their limitations, particularly their underemphasis on long-term structural adaptability and cross-domain interdependencies. The IFR adds to this movement by offering a synthetic but multidimensional framework that policymakers, researchers, and institutions can use to track future preparedness over time and across contexts.

Perhaps most significantly, the IFR introduces a normative shift in how we conceptualize national success. In a world where shocks are no longer rare but routine, and where change is not linear but exponential, success can no longer be measured solely by short-term growth or static indicators of performance. Instead, the new imperative is strategic agility: the ability to anticipate, absorb, and adapt in ways that preserve prosperity and social cohesion amid disruption. Future readiness is not a static end state, but a dynamic capacity that must be cultivated, renewed, and governed with deliberation.

In this light, the IFR is not merely a measurement tool, but a strategic compass. It invites policymakers to reconsider the goals of national development, encourages societies to invest in flexibility and learning, and offers a shared vocabulary through which nations can assess progress not only by where they stand today, but by how prepared they are to meet tomorrow. Future readiness, as articulated here, is not just a condition to be achieved, it is a mindset and a strategy for an age in which disruption is the norm, and thriving requires the courage and the leadership to adapt.

# References

Artadi, Elsa and Xavier Sala-i-Martin, (2004), "The Global Competitiveness Index", World Economic Forum

Bank for International Settlements, (2016), "Economic resilience: a financial perspective" https://www.uschamber.com/assets/archived/images/19_G20+Submission+from+BIS+-+Economic+Resilience_17.pdf

Barro, Robert (1984), "Macroeconomics", Wiley.

Barro, Robert and Xavier Sala-i-Martin, (2004), "Economic Growth", 2nd Edition, MIT Press

Blanchard, Olivier J. and Danny Quah (1989), "The Dynamic Effects of Aggregate Demand and Supply Disturbances", American Economic Review, Vol 79, 4,  pp 655-673.

Briguglio, L., Cordina, G., Farrugia, N., and Vella, S. (2009). Economic vulnerability and resilience: concepts and measurements. Oxford Development Studies, 37(3):229–247.

Brinkmann, Henrik,  Christoph Harendt, Friedrich Heinemann, and Justus Nover (2017), "Economic Resilience: A new concept for policy making?"

Capoani, Luigi , Mila Fantinelli, Luca Giordano (2025), "The concept of resilience in economics: a comprehensive analysis and systematic review of economic literature". Continuity and Resilience Review. https://www.emerald.com/insight/content/doi/10.1108/crr-11-2024-0045/full/html

Cutter, S. L., Barnes, L., Berry, M., Burton, C., Evans, E., Tate, E., and Webb, J. (2008). A place-based model for understanding community resilience to natural disasters. Global environmental change, 18(4):598–606.

European Comission (2012), "Action Plan for Resilience in Crisis Prone Countries: 2013-2020" https://ec.europa.eu/echo/files/policies/resilience/com_2013_227_ap_crisis_prone_countries_en.pdf?utm_source=chatgpt.com

European Comission (2014), "EU Resilience Compendium: Saving lives and livelihoods" https://ec.europa.eu/echo/files/policies/resilience/eu_resilience_compendium_en.pdf?utm_source=chatgpt.com

European Comission (2020), "Strategic Foresight Report", https://commission.europa.eu/strategy-and-policy/strategic-foresight/2020-strategic-foresight-report_en

European Comission (2025), "Resilience Dashboards", https://commission.europa.eu/strategy-and-policy/strategic-foresight/2020-strategic-foresight-report/resilience-dashboards_en

Freightfox (2025), "Supply Chain Disruptions in 2025: Causes, Impact, and Solutions" https://www.freightfox.ai/blog/supply-chain-disruption

Hafele, J., Barth, J., Le Lannou, L-A., Bertram, L., Tripathi, R., Kaufmann, R., Engel, M. (2022): "A framework for economic resilience: guiding economic policy through a social-ecological transition". ZOE Institute for Future-fit Economies: Cologne. https://zoe-institut.de/wp-content/uploads/2023/10/ZOE_Economic_Resilience_Framework.pdf

Hafele, J., Bertram, L., Demitry, N., Le Lannou, L-A., Korinek, L., Barth, J. ( 2023 ): The Economic Resilience Index: assessing the ability of EU economies to thrive in times of change. ZOE Institute for Future-fit Economies: Cologne. https://zoe-institut.de/wp-content/uploads/2023/09/Economic_Resilience_Index_Final-1.pdf

Hallegatte, Stephane (2014), "Economic Resilience : Definition and Measurement." World Bank Policy Research Working Paper No. 6852.  https://openknowledge.worldbank.org/entities/publication/a0878972-07ce-5642-8e8a-ad16d1ec2749

Hardin, Jesse (2025), "In the Aftermath: Understanding Natural Disasters' Impact on the Economy", Equifax Market Trends https://www.equif_ax.com/business/blog/-/insight/article/in-the-aftermath-understanding-natural-disasters-impact-on-the-economy/

Holling C. S. (1973), "Resilience And Stability Of Ecological Systems", Annual Review of Ecology and Systematics Vol. 4.

IMD (2025), "Future Readiness Indicator", https://www.imd.org/future-readiness-indicator/home/latest-rankings

IMD (2025), "World Competitiveness Ranking". https://www.imd.org/centers/wcc/world-competitiveness-center/rankings/world-competitiveness-ranking/

Jawad, Ali Qassim (2020). "Government Reimagined", Thinkers50 Limited, London.

Keynes, John M. (1930), "Economic Possibilities for our Grandchildren", published in "Essays in Persuasion", London: Macmillan 1931.

Kitsos, Tasos (2021), Economic resilience to shocks, University of Bimingham https://research.aston.ac.uk/files/197708011/Economic_resilience_to_shocks_implications_for_labour_markets_-_Feb_21.pdf

Lucas, Robert E., (1977), "Understanding Business Cycles", Carnegie Rochester Conference Series on Public Policy. https://www.sciencedirect.com/science/article/abs/pii/0167223177900021

Mester, Loreta (2024), "Building Financial System Resilience", Federal Reserve Bank of Cleveland

Nelson, Charles R. and Charles I. Plosser, (1982), "Trends and Random Walks in macroeconòmics time series", Journal of Monetary Economics, 10, pp 139-162.

OECD (2008), Handbook on Constructing Composite Indicators: Methdology and USer Guide. *OECD Journal: Economic Studies, Volume 2008*, OECD Publishing, Paris, https://www.oecd.org/en/publications/handbook-on-constructing-composite-indicators-methodology-and-user-guide_9789264043466-en.html

OECD Joint Research Centre-European Commission (2008), "Handbook on constructing composite indicators: methodology and user guide". OECD publishing. https://www.oecd.org/en/publications/handbook-on-constructing-composite-indicators-methodology-and-user-guide_9789264043466-en.html

Reggiani, Aura et al (2001), "Resilience: an evolutionary approach to special systems", Tirbergen Institute Discussion Paper

Rezai, Armon, Franz Ruch, Rishabh Choudhary, John Francois, and Suhyun Lee (2025), "Can fiscal policy help countries face extreme-weather shocks?", Prevention Web https://www.preventionweb.net/news/can-fiscal-policy-help-countries-face-extreme-weather-shocks

Rose, A. (2004). Defining and measuring economic resilience to disasters. Disaster Prevention and Management: An International Journal, 13(4):307–314.

Rose, A. (2007). Economic resilience to natural and man-made disasters: Multidisciplinary origins and contextual dimensions. Environmental Hazards, 7(4):383–398.

Rose, A. Z. (2009). Economic resilience to disasters. Technical Report Research report 8, The Community and Regional Resilience Institute.

Smith, Tyler (2025), "Climate shocks and the economy. Is the US economy adapting to more severe weather patterns?", Research Highlights of the American Economic Association https://www.aeaweb.org/research/severe-weather-macroeconomy-us

Taleb, Nassim (2012), "Antifragile", Penguin Random House

Tol, Richard S. (2018),"The Economic Impacts of Climate Change", Review of Environmental Economics and Policy

U.S. Bank (2025), How to Handle Market Volatility. Wealth Management Division. https://www.usbank.com/investing/financial-perspectives/investing-insights/how-to-handle-market-volatility.html

U.S. Development Administration, (2025), "Economic Resilience". https://www.eda.gov/sites/default/files/2025-02/2025_CEDS_Content_Guidelines.pdf

Vugrin, E., Warren, D., Ehlen, M., Camphouse, R., (2010), "A Framework for Assessing the Resilience of Infrastructure and Economic Systems", in: Gopalakrishnan, K., Peeta, S. (Eds.), Sustainable and Resilient Critical Infrastructure Systems: Simulation, Modeling, and Intelligent Engineering. Springer, Berlin, Heidelberg, pp. 77–116.

World Economic Forum (2025), "The Global Risks Report". https://www.weforum.org/publications/global-risks-report-2025/digest/

World Intellectual Property Organization (WIPO), (2025), "The Global Innovation Index". https://www.wipo.int/en/web/global-innovation-index

Yu, Howard, Jialu Shan, Angelo Boutalikakis, Lawrence Tempel and Zuriati Balian (2022), "What Makes a Company "Future Ready?", Harvard Business Review, March. https://hbr.org/2022/03/what-makes-a-company-future-ready

Zehra, Taheer (2025), "Ranking Countries by Labour Market Flexibility in 2025", qreos.